\documentclass[12pt]{article}
\usepackage{amsmath, amssymb,natbib}
\usepackage{graphicx,psfrag,epsf}
\usepackage{enumerate}
\usepackage{graphics,pstricks,epstopdf}
\usepackage[utf8]{inputenc}
\usepackage{subfigure}
\usepackage{epsfig}
\usepackage{subfigure}
\usepackage{txfonts}
\usepackage{palatino}
\usepackage[figuresright]{rotating}
\usepackage{bm}
\usepackage{ragged2e}
\usepackage{enumitem}
\usepackage{pgfplots}
\usepackage{tikz}
\usepackage{tkz-tab}
\usepackage{tabto}

\usepackage{lscape}

\usepackage{hyperref}
\usepackage{algorithm}
\usepackage{algorithmic}
\usepackage{setspace}
\newtheorem{thm}{Proposition}
\newcommand*{\QED}{\hfill\ensuremath{\blacksquare}}

\newcommand{\blind}{0}

\pgfmathdeclarefunction{gauss}{3}{%
  \pgfmathparse{#3*1/(#2*sqrt(2*pi))*exp(-((x-#1)^2)/(2*#2^2))}%
}

\begin{document}

\def\spacingset#1{\renewcommand{\baselinestretch}%
{#1}\small\normalsize} \spacingset{1}


\if0\blind
{
  \title{\bf A Scalable Empirical Bayes Approach to Variable
Selection in  Generalized Linear Models}
  \author{Haim Y. Bar\thanks{
    The authors gratefully acknowledge the following funding support: Prof. Bar's research was supported by NSF-DMS
1612625. Professor Booth's research was partially supported
by an NSF grant, NSF-DMS 1208488, and by NSF-DMS 1611893. Professor Wells' research was partially supported by
NSF-DMS 1208488, NSF-DMS 1611893, and NIH grant U19 AI111143.}\hspace{.2cm}\\
    Department of Statistics University of Connecticut,\\ Storrs CT, USA.\\
    and \\
    James G. Booth and Martin T. Wells\\
    Department of Statistics and Data Science, Cornell University, \\ Ithaca NY, 14853, USA.}
  \maketitle
} \fi

\if1\blind
{
  \bigskip
  \bigskip
  \bigskip
  \begin{center}
    {\LARGE\bf A Scalable Empirical Bayes Approach to Variable
Selection in  Generalized Linear Models}
\end{center}
  \medskip
} \fi

\bigskip

\begin{abstract}
A new empirical Bayes approach to variable selection in the context of
generalized linear models is developed. The proposed algorithm scales to
situations in which the number of putative explanatory variables is very large, possibly
much larger than the number of responses. The coefficients
in the linear predictor are modeled as a three-component mixture
allowing the explanatory variables to have a random positive effect on the response, a
random negative effect, or no effect. A key assumption is that only a
small (but unknown) fraction of the candidate variables have a
non-zero effect. This assumption, in addition to treating the
coefficients as random effects facilitates an approach that is computationally
efficient. In particular, the number of parameters that have to be estimated is small, and
remains constant regardless of the number of explanatory variables.
The model parameters are estimated using a Generalized Alternating Maximization
algorithm which
is scalable, and leads to significantly faster convergence compared
with simulation-based fully Bayesian methods.
\end{abstract}

\noindent%
\vfill

\newpage
\spacingset{1.45} 

\section{Introduction}
This paper concerns variable selection in generalized linear regression models
when there are a large number of candidate explanatory variables
(putative predictors), most of which
have little or no effect on the dependent variable. An empirical
Bayes, model-based approach is proposed that is implemented via a fast
and scalable Generalized Alternating Maximization algorithm.

The new age of high speed computing and technological advances in
genetics and molecular biology, for example, have dramatically changed
modeling and computation needs. It is now common for researchers
to want to estimate the effects of hundreds or
even thousands of predictors ($K$) on a given response, often with a far
smaller sample size ($N$).  In such cases, traditional fitting methods such
as least squares break down. In addition, even with a relatively modest
number of predictors, the model space can be large enough to render exhaustive
search-based algorithms impractical.

Automated methods for variable selection in normal linear regression models have
long been studied in the literature; see, for example
\cite{Hocking:1976}. Nowadays virtually every
statistical package contains an implementation of standard stepwise
methods that typically add or remove one variable from the model in each
iteration, based on sequential F-tests and a threshold, or a well-known
selection criterion such as AIC , BIC, or Mallow's-$C_p$.
A modern alternative is to use false discovery rate (FDR) for
stepwise model selection \citep{benjamini2009}.

Much of the recent literature has focused on variations of penalized likelihood approaches
in which coefficient estimation and variable selection are done simultaneously.
The most well-known method of this type is the LASSO \citep{Tibshirani:1996}
which minimizes the residuals sum of squares subject to an $\ell_1$ constraint.
This constraint allows the number of non-zero parameter
estimates to be controlled and adapt to sparsity.
Other related methods that are based on a minimizing
a loss function, subject to a constraint on the complexity of the
model, include SCAD \citep{FanLi:2001}, the adaptive LASSO \citep{Zou:2006},
LARS \citep{Efron:2004,Hesterberg:2010}, and more recent proposals by
\cite{Malgorzata2015}, \cite{buhlmann2014} and \cite{LedererM15}.

Bayesian approaches are another important direction in model selection
research. Significant contributions include \cite{george:1993},
\cite{casella:2006}, and the spike-and-slab method in
\cite{Ishwaran:2005}.  The model proposed here is similar to
\cite{zhang:2005}, and \cite{Guan:2012} whose work is motivated by QTL
and genome-wide association studies (GWAS).  Our model allows for a
fully-Bayesian implementation, but an empirical Bayes analysis via the
Generalized Alternating Maximization algorithm \citep{Gunawardana:2005} is
proposed instead because the running time of an MCMC sampler is too
long for many data sets in modern applications.
For example, in our simulations (Section \ref{sec:sim}) we find that the faster of
two MCMC-based variable selection methods required  40 minutes to
complete 1000 MCMC iterations. In contrast, non-MCMC methods required
only a few seconds to perform variable selection, and generally, gave better results.
In this sense the
algorithm is a close competitor to the exact EM algorithm of
\cite{Rockova:2014} based on a Bayesian spike-and-slab model. A key
advantage of the approach proposed in this paper is that it extends in a straightforward
manner to the generalized linear model framework.

Recently developed continuous prior distributions have proven more effective in sparse regression than the a Bayesian lasso \citep{park2008bayesian} that uses a Laplace prior, which fails to simultaneously induce sparsity while efficiently recovering non-null parameters \citep{van2016conditions}. Among these are “global-local priors” including the Horseshoe-type priors \citep{carvalho2010horseshoe, bhadra2017horseshoe+} and the Gamma Gamma prior \citep{bai2017inverse}. These approaches offer computational advantages relative to two-component spike-and-slab mixture priors since one does not need to explore a complex discrete model space of size $2^K$. A limitation of the continuous shrinkage prior approach is that it fails to provide a sparse solution. To address the problem, ad hoc post-processing methods for producing sparse estimates from posterior samples that decoupled shrinkage and selection   have been developed \citep{hahn2015decoupling}.

The remainder of the article is organized as follows.  The model and
notation are introduced in Section \ref{sec:vs:model}.  A Generalized Alternating Maximization
fitting algorithm and selection procedure is described
in Section \ref{sec:vs:est}. Section \ref{sec:sim} describes results
of simulation studies in which the proposed procedure is compared with
several variable selection packages, including
\texttt{ncvreg} \citep{Rncvreg}
and \texttt{SIS} \citep{SIS2018}, both of which implement three types of penalties:
LASSO \citep{Tibshirani:1996}, SCAD \citep{FanLi:2001}, and MCP
\citep{Zhang:2010}. Other packages we included in our simulations are: \texttt{glmnet} \citep{glmnet},
\texttt{lars} \citep{lars2013}, \texttt{EMVS}  \citep{EMVS2018},
\texttt{spikeslab} \citep{spikeslab2013}, \texttt{mombf} \citep{mombf2018}, and
\texttt{BoomSpikeSlab} \citep{Scott2017}. We also compared the performance of our method
with a one-predictor-at-a-time approach, controlling the false discovery rate \citep{benjamini:hochberg}.
In Section \ref{sec:vs:cases} we discuss some applications. First,
we demonstrate our method when the response is normal (the logarithm of vitamin B12
production rate). In this case the sample size is $N=71$ and the number of predictors is 4,088.
We compare the model obtained from our method with ones obtained from TREX \citep{LedererM15},
and the results obtained by \cite{buhlmann2014}. The second example demonstrates an
application to a binary response (categorized Body Mass Index), which we compare to an analysis
using the continuous outcome.
In this case, $N=96$ and there are
45 (compositional) predictors. Finally, we explain how to perform variable selection
in survival analysis by treating the number of deaths in a sequence of
non-overlapping time intervals as a Poisson counts.
The paper concludes with with some discussion in Section \ref{sec:varsel:conc}.
Important implementation considerations and additional examples are
discussed in the supplementary material.

\section{A Statistical Model for Automatic Variable
Selection}\label{sec:vs:model}
Consider  responses $y_i$, $i=1,\ldots,N$, and assume that the
mean of the $i$th response, $\lambda_i$, is linked to a linear predictor, $\eta_i$, as follows:
\begin{eqnarray}\label{model_1}
  g(\lambda_{i})\equiv\eta_i=\sum_{j=1}^{J}x_{ij}\beta _{j}+
  \sum_{k=1}^{K}z_{ik}\gamma _{k}u_{k}\,.
\end{eqnarray}
The model (\ref{model_1}) allows for $J\ge 0$ predictors, $x_{ij}$, that are
always included in the model and a set of $K>0$ `putative' predictors, $z_{ik}$,
from which it is expected only a small subset are to be
included. Here, $\beta_j$ is the coefficient associated with the $j$th
'locked-in' predictor, $u_k$ is a random coefficient associated with
the $k$th putative predictor. We assume that each $\gamma _{k}u_{k}$
belongs to one of three components,  $C_L$, $C_0$, and $C_R$,
such that $\gamma _{k}=0$ for $k\in C_0$,  $\gamma _{k}=-1$ for $k\in C_L$,
and $\gamma _{k}=1$ for $k\in C_R$.
We also assume that $u_{k} \stackrel{iid}{\sim} N\left( \mu ,\sigma ^{2}\right)$,
independently of $\gamma _{k}$.
Justification for both of these choices is given in
Section~\ref{sec:vs:est}.
Thus, the inclusion of the $k$th predictor in the linear model is determined by
the value of $\gamma _{k} \stackrel{iid}{\sim} multinomial\left(-1,0,1;
  p_L,p_0,p_R\right)$.
Specification of the distribution of the responses is completed by
assuming that, conditional on the linear predictors, they are
independent draws from a particular exponential dispersion family,
with the most important special cases being the normal, binomial and
Poisson distributions.
Within this modeling framework the problem of
variable selection is cast as a classification problem, for which the main
interest lies in identifying which putative variables belong to $C_L\cup C_R$,
i.e., which latent variables, $\gamma_k$, are non-zero.

Let $\mathbf{X}$ denote the $N\times J$ matrix with
$j$th column, $\mathbf{x}_{j}$, containing the values of the $j$th
`locked in' predictor. Similarly, define $\mathbf{Z}$ to be the
$N\times K$ matrix with $k$th column $\mathbf{z}_{k}$, the
corresponding vector for the $k$th putative predictor.  Then the
linear predictor model (\ref{model_1}) can be rewritten in matrix form
as
\begin{eqnarray*}
  \bm{\eta}=\mathbf{X}\bm{\beta}+\mathbf{Z}\bm{\Gamma}\mathbf{u}
\end{eqnarray*}
where $\bm{\beta} =(\beta_1,...,\beta_J)'$, $\bm{\Gamma} \equiv diag\left( \gamma _{1},\gamma
_{2},\ldots,\gamma _{K}\right)$. Furthermore, the distributional assumptions
concerning $\{u_k\}$ imply
\begin{eqnarray*}
  \mathbf{Z}\bm{\Gamma}\mathbf{u}\,|\,\bm{\Gamma} &\sim N\left(
\mathbf{Z}\bm{\Gamma}\bm{\mu},\sigma ^{2}\mathbf{Z}\bm{\Gamma}^2\mathbf{Z}'\right)
\end{eqnarray*}
where $\bm{\mu}=\mathbf{1}_K\mu$.

Mixture models provide simplicity and tractability and are very
popular in many applications. However, they are known to have
undesirable mathematical properties, such as unbounded likelihood and
lack of identifiability \citep{chen2009,McLachlan2000}. For example, in the
Gaussian case, if $\mu=0$ then the product $\gamma_ku_k$ in
model~(\ref{model_1}) has marginal density of the form
$h(x)=p_0\cdot 0+(p_L+p_R)\varphi(x;0,\sigma)$, where $\varphi(\cdot;\mu,\sigma)$
denotes a normal density. Clearly, $p_L$ and $p_R$ cannot
be separately identified in such a model. However, when $\mu\not=0$,
the density is
$h(x)=p_0\cdot 0 +
p_L\varphi(x;-\mu,\sigma)+p_R\varphi(x;\mu,\sigma)$,
which is identifiable if $p_L\not= p_R$. Moreover, even if
$p_L=p_R$, the identifiability issue only concerns the sign of
$\mu$ which, in this case, does not affect the marginal distribution $h$.

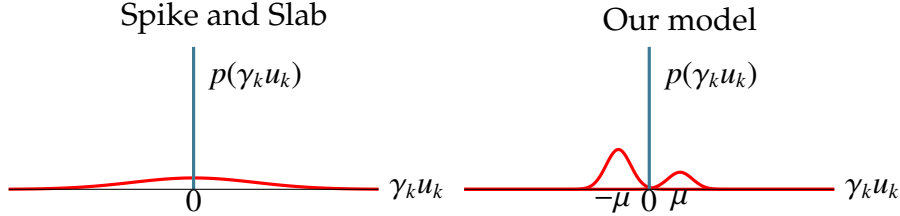
\begin{figure}
\centering
\begin{tikzpicture}[baseline=0pt]
     \begin{axis}[
  no markers, domain=-3:3, samples=1000,
  hide y axis,
  axis lines*=left, xlabel=$\gamma_k u_k$, ylabel=$p(\gamma_k u_k)$,
  every axis x label/.style={at=(current axis.right of origin),anchor=west},
  height=4cm, width=6.5cm,
  xtick={4,6.5}, ytick=\empty,
  enlargelimits=false, clip=false, axis on top,
  grid = major
  ]
  \addplot [very thick,white] {gauss(0,0.1,0.8)};
  \addplot [very thick,red] {gauss(0,1,0.5)};
  \draw [very thick,cyan!50!black] (300,0) -- (300, 250);
  \node at (400,200) {$p(\gamma_k u_k)$};
  \node at (300,-20) {$0$};
  \node at (350,300) {Spike and Slab};
\end{axis}
\end{tikzpicture} \begin{tikzpicture}[baseline=0pt]
     \begin{axis}[
  no markers, domain=-3:3, samples=1000,
  hide y axis,
  axis lines*=left, xlabel=$\gamma_k u_k$, ylabel=$p(\gamma_k u_k)$,
  every axis x label/.style={at=(current axis.right of origin),anchor=west},
  height=4cm, width=6.5cm,
  xtick={4,6.5}, ytick=\empty,
  enlargelimits=false, clip=false, axis on top,
  grid = major
  ]
  \addplot [very thick,white] {gauss(0,0.1,0.8)};
  \addplot [very thick,red] {gauss(0.5,0.2,0.15)};
  \addplot [very thick,red] {gauss(-0.5,0.2,0.35)};
  \draw [very thick,cyan!50!black] (300,0) -- (300, 250);
  \node at (400,200) {$p(\gamma_k u_k)$};
  \node at (350,300) {Our model};
  \node at (300,-20) {$0$};
  \node at (350,-20) {$\mu$};
  \node at (240,-20) {$-\mu$};
\end{axis}
\end{tikzpicture}
\caption{A graphical representation of the spike and slab model vs. our mixture model.}\label{modelgraphs}
\end{figure}

Our hierarchical mixture model is similar to the well-known Spike and Slab
model of \cite{Ishwaran:2005} which is implemented, for example, in the R
package BoomSpikeSlab \citep{Scott2017} and in EMVS \citep{Rockova:2014}.
The main difference is our choice of a three-way mixture model, in which
there are two non-null components, rather than one.
Compared with the spike and slab approach, our model  offers a couple of advantages.
To illustrate these advantages, it is helpful to plot the theoretical distributions
of $\gamma_k u_k$ under the two models (Fig. \ref{modelgraphs}). The two-component mixture
is depicted on the left, and our model on the right. In both cases, $p_0=0.8$.

\vspace{3 mm}

The two-component mixture assumes that the non-null distribution
is symmetric, which implies a prior belief that the proportion of variables which are
positively correlated with the response is the same as the proportion
of predictors which are negatively correlated with the response. This may be an unreasonable
assumption, and the three-component mixture model is more flexible in
this regard.

In addition, the non-null component in the two-component mixture has much of its
mass around zero, which is counterintuitive because it is assumed that
variables in the non-null component have a non-zero effect. In
contrast, the three-component model assigns a very small probability
to non-null values near zero. Our mixture model also allows for the non-null components to
be highly concentrated, which may be especially useful in situation where there is
a single significant predictor.
The assumption of symmetry of the two nonnull components in our model may be relaxed, but it has
some important benefits. First, it implies that the model is invariant to the sign of columns
of $\mathbf{Z}$. Second, it allows us to borrow information across the two nonnull
components, which greatly contributes to computational stability, as well as power to detect true predictors,
and maintaining a low false positive rate. This is especially beneficial when at least one of the
nonnull components  consists of a small number of  predictors.

The three component prior can be viewed as similar to a non-local prior for which density functions are identically zero whenever a model parameter is equal to its null value.  Conversely, spike and slab priors are local priors where component densities are positive at null parameter values.  \cite{johnson2010, johnson2012} demonstrate that model selection procedures based on non-local prior densities assign a posterior probability of one to the true model as the sample size $n$ increases and certain regularity conditions on the design
matrix pertain.  Furthermore under the same conditions, they show that standard approaches based on local prior specifications result in the asymptotic assignment of a posterior probability of zero to the true model.


\section{Estimation and Variable Selection}\label{sec:vs:est}
\subsection{The Complete Data Likelihood}
The marginal distribution of the responses according to the model
described in the previous section is determined by the
parameter vector $\bm{\theta}=\{\bm{\beta}, \mu, \sigma^2, \bm{p}, \phi\}$,
where $\phi$ is a dispersion parameter which, depending on the
GLM specification, may or may not be known.  Define
$n_j=\sum_{k=1}^KI(\gamma_k=j)$, for $j=-1,0,1$ corresponding to the left,
middle, and right components of the mixture model, respectively, and let
$\varphi(\cdot;\bm{\mu},\bm{\Sigma})$ denote a multivariate normal
density with mean $\bm{\mu}$ and covariance matrix $\bm{\Sigma}$. Then assuming
a canonical link function, the complete data log likelihood
is
\begin{eqnarray}
  \label{cloglik}
  \ell(\mathbf{y},  \bm\gamma| \bm\theta)=\log f_C(\mathbf{y}, \bm{\gamma} | \bm\theta)
  &=& \log\int\exp\left\{
  \sum_{i=1}^N\frac{m_i}{\phi}[\eta_iy_i-b(\eta_i)]
      \right\}\varphi\left(\mathbf{u};\bm{\Gamma}\bm{\mu},\sigma^2\mathbf{Z}\bm{\Gamma}^2\mathbf{Z}'\right)d\mathbf{u}
      \nonumber \\
  && +\sum_{i=1}^Nc(y_i,\phi/m_i) +\sum_{j=-1,0,1}n_j\log p_j\,,
\end{eqnarray}
where the $m_i$'s are known positive weights, $b$ is the cumulant generator for the GLM satisfying
$b'(\eta_i)=\lambda_i$, and $b''(\eta_i)= V(\lambda_i)$, where $V$ is the GLM variance function.

In the Gaussian case the integral in (\ref{cloglik}) has a closed form
and the complete data loglikelihood reduces to
\begin{eqnarray}
  \label{gaussian-cloglik}
    \ell(\mathbf{y},  \bm\gamma| \bm\theta)
 \!\! \! \! &=& \! \! \! \! -\frac{1}{2}
  (\mathbf{y}-\mathbf{X}\bm{\beta}-\mathbf{Z}\bm{\Gamma}\bm{\mu})'
  (\phi\mathbf{W}^{-1}+\sigma^2\mathbf{Z}\bm{\Gamma}^2\mathbf{Z}')^{-1}
  (\mathbf{y}-\mathbf{X}\bm{\beta}-\mathbf{Z}\bm{\Gamma}\bm{\mu})
                                       \nonumber \\
  && \! \!\! \!  -\frac{1}{2}\log|\phi\mathbf{W}^{-1}\!+\sigma^2\mathbf{Z}\bm{\Gamma}^2\mathbf{Z}'| +
   \! \! \! \!  \sum_{j=-1,0,1}\! \!n_j\log p_j-\frac{N\log(2\pi)}{2},
\end{eqnarray}
where $\mathbf{W}=\mbox{diag}(w_i)$ and $w_i\equiv m_i$.
For non-Gaussian GLMs an approximate complete data log-likelihood is
obtained by substituting the so-called 'working response' and
'iterative weight matrix' in place of $\mathbf{y}$ and $\mathbf{W}$ in (\ref{gaussian-cloglik}), with
components given by
\begin{eqnarray*}
  \tilde{y}_i=g(\tilde\lambda_i)+g'(\tilde\lambda_i)(y_i-\tilde\lambda_i)
  \hspace{3mm}\mbox{and}\hspace{3mm}
  \tilde{w}_i=m_i/g'(\tilde\lambda_i)\,.
\end{eqnarray*}
The substitutions can be justified on
the basis of a Laplace approximation to the integral in
(\ref{cloglik}), and are the basis of numerous algorithms
in the literature for fitting GLMs with random effects. See, for
example, \cite{scha:1991,bres:clay:1993,wolf:ocon:1993,mcgi:1994}.

\subsection{A Generalized Alternating Maximization Algorithm}\label{sec:gam}
In principle, 
the ML estimate of the model parameter vector,
$\bm{\theta}$, can be obtained using the EM algorithm
\citep{Dempster1977} using the complete data log-likelihood given in
(\ref{gaussian-cloglik}),
with the Q-function given by
$Q(\bm\theta;\bm\theta') = E_{\bm\theta'}\left\{ \log
    f_c(\mathbf{y},\bm\gamma|\bm\theta)\,|\,\mathbf{y} \right\}$ where $\bm\theta'$
 denotes the current estimate of $\bm\theta$.
However, in this case the expectation is intractable, so we propose using the simple plug-in approximation
\begin{eqnarray}
  \label{E-step}
  E_{\bm\theta'}\left\{ \log
    f_c(\mathbf{y},\bm\gamma|\bm\theta)\,|\,\mathbf{y} \right\} \approx E_{\bm\theta'}\left\{ \log
    f_c(\mathbf{y},\bm\gamma'|\bm\theta)\,|\,\mathbf{y} \right\}\,,
\end{eqnarray}
where $\bm{\gamma}'$ is obtained using one of the methods described later in
this section.

Since the expectation terms are approximated, the resulting iterative estimation procedure
will not fall within the EM framework, or even within the GEM framework \citep{wu1983}, and thus,
convergence results from these frameworks will not apply.
This type of EM variant, where the E-step is replaced with an approximation is an example of
the Generalized Alternating Minimization (GAM) framework of \cite{Gunawardana:2005}, of which
EM and GEM are special cases. We will use the GAM theory to show the convergence of our algorithm,
which we describe in Algorithm \ref{alg1}.

\begin{algorithm}
\setstretch{1}
\caption{The alternating maximization algorithm for fitting model (\ref{model_1})}
\label{alg1}
\begin{algorithmic}[1]
\STATE Initialize $\bm\gamma'$, and choose $\delta\ge 0$\label{inits}
\LOOP
\STATE $\bm\theta' \leftarrow \arg\max_{\bm\theta} \ell(\mathbf{y}, \bm\theta~|~  \bm\gamma')$ \label{maxtheta}
\STATE Calculate $\ell' \leftarrow \ell(\mathbf{y}~|~ \bm\theta', \bm\gamma')$
\FOR{$k=1$ to $K$}
\FOR{$j \in \{-1,0,1\}$}
\STATE Let $\bm\gamma^{*} \leftarrow \bm{\gamma}'$ and set only the $k$-th component $\gamma^{*}_k \leftarrow j$
\STATE Calculate $d_{j,k} \leftarrow \ell(\mathbf{y}| \bm\theta', \bm\gamma^{*}) - \ell'$
\ENDFOR
\ENDFOR
\STATE $S\leftarrow \left\{k\,:\,d_{j,k}>\delta,\text{ for some } j\in\{-1,0,1\}\right\}$
\IF{$S = \emptyset$}
\STATE The algorithm terminates.
\ELSE
\STATE Choose $k\in S$ and update $\gamma'_k$ (leaving all other components in $\bm\gamma'$ unchanged)\label{choosing_k}
\ENDIF
\ENDLOOP
\RETURN $\bm\gamma'$
\end{algorithmic}
\end{algorithm}

This algorithm uses a 'likelihood-ratio' approach because a latent variable $\gamma_k$ is changed from its
current value if that change increases the likelihood in a meaningful way
(holding all other latent indicators at their current values).
The set $S$ consists of
all the variables which yield an improvement greater than  $\delta$ in the loglikelihood when their
current classification according to the three-component mixture model is
changed, while holding all other $\gamma_k$s at their current values.
If $S$ is not empty, choosing $k$ from this set in Line \ref{choosing_k}
is done according to one of the following methods:
\begin{itemize}[noitemsep,topsep=0pt,leftmargin=0.25in]
 \item \textbf{Greedy}: choose $k\in S$ for which $d_{j,k}$ is largest.
 \item \textbf{Weighted probability}: choose $k\in S$ with probability
 $$\frac{d_{j,k}}{\sum_{(r,s)\in S}d_{r,s}}\,.$$
\end{itemize}

\begin{thm}\label{convergencethm}
For the model in (\ref{cloglik}) Algorithm \ref{alg1} converges to a stationary point in a finite number of steps.
\end{thm}
A proof of the proposition is given in Appendix \ref{proof}.

\subsection{Notes and Further Details}
First, recall that in the binomial and Poisson models we work with the `working response and weights',
so $\tilde{y}_i$ and $\tilde{w}_i$ have to be iteratively updated at each iteration of Algorithm \ref{alg1}.

Second, we emphasize that our algorithm modifies at most one coordinate in each iteration because,
changing more than one variable
may introduce multicollinearity. What this means is that even though each single variable
may increase the likelihood, changing a set of variables which are highly correlated
may cause the log-likelihood to decrease (and it may even approach $-\infty$, because $\ell$
involves the logarithm of the precision matrix.)
As a consequence, our algorithm automatically prevents selecting models with highly correlated
predictors. We do, however, keep track of predictors that are correlated with ones selected to be
in the model, since they are likely to be related to the outcome as well. We return to this point in the
Case Studies section.


Third, we may choose $\bm\gamma'$ in Line \ref{inits} to be the variables selected by any other
method (FDR, SIS, EMVS, etc.)
Then, the log-likelihood of the final model selected by our algorithm will be greater than or equal to
the one obtained at the $0^{th}$ iteration.

Fourth, as a referee pointed out, the parameter space for $(\bm\theta, \bm\gamma)$ cannot assumed to be unimodal,
especially when some putative variables are correlated with each other or with columns in  $\mathbf{X}$. Therefore, it
is recommended to run Algorithm \ref{alg1} multiple times, using the
weighted probability approach for selecting the next putative variable to be updated.
This approach is feasible because our method is computationally efficient. When the posterior distribution
is multimodal there is no one `correct' model, and running the algorithm multiple times will allow users to obtain
different, but possibly equally relevant sets of significant predictors each time they fit the model.
This approach is demonstrated in Section \ref{sec:vs:cases} (the riboflavin data example).

Finally, to obtain the maximum likelihood estimates $\bm\theta'$ in Line \ref{maxtheta},
the following formulas are used.
Let $\bm{\Sigma}=\phi\mathbf{W}^{-1}+\sigma^2\mathbf{Z}\bm{\Gamma}^2\mathbf{Z}'$
and $\mathbf{H}=[\mathbf{X},\mathbf{Z}\bm{\Gamma}\mathbf{1}]$, the update
formula for $\tilde{\bm{\beta}}=(\bm{\beta}',\mu)'$ is given by
\begin{eqnarray*}
\tilde{\bm{\beta}}=(\mathbf{H}'\bm{\Sigma}^{-1}\mathbf{H})^{-1}
\mathbf{H}' \bm{\Sigma}^{-1}\mathbf{y}\,
\end{eqnarray*}
so $\mu$ is simply the last element in $\tilde{\bm{\beta}}$,
and the updates for the variance components are given by
\begin{eqnarray*}
\phi=\frac{\tau_e}{N}\label{sigma_e}
\hspace{3mm}\mbox{and}\hspace{3mm}
\sigma^2=\frac{\tau_r}{\mbox{rank}(\mathbf{Z}\bm{\Gamma})}\,
\end{eqnarray*}
provided $\mbox{rank}(\mathbf{Z}\bm{\Gamma})>0$ and $\mu=\sigma^2=0$ otherwise,
where
\begin{eqnarray*}
\tau_e&=&\mbox{trace}[\phi\mathbf{I}_N-\phi^2\bm{\Sigma}^{-1}]+
\phi^2(\mathbf{y}-\mathbf{H}\bm{\tilde{\beta}})'\bm{\Sigma}^{-2}
(\mathbf{y}-\mathbf{H}\bm{\tilde{\beta}})\\
\tau_r&=&\mbox{trace}[\sigma^2\mathbf{I}_K-\sigma^4\bm{\Gamma}\mathbf{Z}'
\bm{\Sigma}^{-1}\mathbf{Z}\bm{\Gamma}]+
\sigma^4(\mathbf{y}-\mathbf{H}\bm{\tilde{\beta}})' \bm{\Sigma}^{-1}
\mathbf{Z}\bm{\Gamma}^2\mathbf{Z}'
\bm{\Sigma}^{-1}(\mathbf{y}-\mathbf{H}\bm{\tilde{\beta}})\,.
\end{eqnarray*}
(see Section 8.3.b in \citealt{McCulloch:1992}).
Finally, maximizing (\ref{gaussian-cloglik}) with respect
$p_L,p_{0},p_R$ leads to the updates $p_{j}=n_j/K$ for $j=-1,0,1$.

Small values of $\mu$ are counterintuitive because they suggest that the mean effects of
selected putative variables are close to zero. Moreover, as noted
earlier, small values of $\mu$ can lead to identifiability problems.
To prevent such problems one might consider adding a penalty term on
$\mu$ as proposed by \cite{chen2009}.
Another possible approach to preventing identifiability problems is to use a different
nonnull prior, such as a mixture of two log-normal distributions,
so that the probability that variables with effect size close to zero will be considered as
nonnull, will be practically 0.
However, as we shall see in the next subsection, one of the strengths of our method is
that it allows us to achieve significant dimension reduction via the Woodbury identity, which requires the
normality assumption. So, were we to use a different nonnull distribution, we would need to add an
extra step to the algorithm in order to normalize the nonnull components (via a Laplace approximation, for example).
In practice, however, we find that identifiability problems
are effectively avoided by choosing appropriate initial values
for our Generalized Alternating Maximization algorithm.
For further details regarding the initialization of the algorithm, see the
Supplementary Material.

%
%

\subsection{Modifications for large $N$ and $K$}
The complete data
log-likelihood (\ref{cloglik}) contains a large ($N\times N$) matrix which has to be inverted
to compute the iterative approximate ML estimates.
However, using the Woodbury identity \citep{Golub:1996},
\begin{eqnarray*}
  \bm{\Sigma}^{-1}\!\!& = & \left(\phi\mathbf{W}^{-1}+\sigma^2\mathbf{Z}\bm{\Gamma}^2\mathbf{Z}'\right)^{-1} \!\! \! \!  \!\!  \\
   &=&  \!\! \! \!
  \frac{1}{\phi}\left[\mathbf{W}^{\frac{1}{2}}\left(\mathbf{I}_N+\frac{\sigma^2}{\phi}\mathbf{W}^{\frac{1}{2}}\mathbf{Z}\bm{\Gamma}^2\mathbf{Z}'\mathbf{W}^{\frac{1}{2}}\right)^{-1}\! \!\mathbf{W}^{\frac{1}{2}}\right] \\
&=& \!\! \! \! \frac{1}{\phi}\left[\mathbf{W}\!- \! \frac{\sigma^2}{\phi}\mathbf{W}^{\frac{1}{2}}\mathbf{Z}\bm{\Gamma}'
\left(\mathbf{I}_K \!+\!\frac{\sigma^2}{\phi}\bm{\Gamma}'\mathbf{Z}'\mathbf{W}\mathbf{Z}\bm{\Gamma}\right)^{-1} \!\!   \bm{\Gamma}'\mathbf{Z}'\mathbf{W}^{\frac{1}{2}}\right].
\end{eqnarray*}
This simplifies the computations considerably because the $(k,l)th$ element of
$\bm{\Gamma}'\mathbf{Z}'\mathbf{W}\mathbf{Z}\bm{\Gamma}$ is
proportional to $\gamma_k\gamma_l$. Specifically, suppose there are $L$ variables for which
$\gamma_k\neq 0$. Define
$\bm{\Gamma}_L$ to be the $L\times L$ reduced matrix in
which rows and columns of $\bm{\Gamma}$ corresponding to excluded putative variables
have been eliminated. Similarly, define $\mathbf{Z}_L$ by eliminating
the corresponding rows of $\mathbf{Z}$. Then
\begin{eqnarray*}
  \bm{\Sigma}^{-1}=\frac{1}{\phi}\left[\mathbf{W}-\frac{\sigma^2}{\phi}\mathbf{W}^{1/2}\mathbf{Z}_L\bm{\Gamma}'_L
\left(\mathbf{I}_L+\frac{\sigma^2}{\phi}\bm{\Gamma}'_L\mathbf{Z}'_L\mathbf{W}\mathbf{Z}_L\bm{\Gamma}_L\right)^{-1}\bm{\Gamma}'_L\mathbf{Z}'_L\mathbf{W}^{\frac{1}{2}}\right]\,.
\end{eqnarray*}
Thus, inversion of the $N\times N$ matrix $\bm{\Sigma}$ is reduced to
inverting a much lower dimensional $L\times L$ matrix. Similar
simplifications, due to the exclusion of most putative variables, apply
to computation of the determinant term in (\ref{gaussian-cloglik}).

Further details about implementation considerations are provided in the Supplementary
Materials. Specifically, we discuss how to deal with correlation and
interactions among the putative variables, and we discuss computational challenges
stemming from the large number of variables, such as how to avoid loading
the entire matrix $\mathbf{Z}$ to memory, and how to parallelize the algorithm.

\section{Simulations}\label{sec:sim}
We conducted a simulation study to verify that under the assumed model (\ref{model_1})
the algorithm yields accurate parameter estimates (not shown here), and to compare the performance of our
algorithm in terms of power and accuracy with other methods.
Not surprisingly, as the sample size increases (even when $N$ is still much smaller than
$K$), the parameter estimates become more accurate, the power to detect the non-null
variables increases, and the Type-I error rate decreases.

In this section we focus on the results of a simulation study that compares the
Type I and Type II errors when the data are not  necessarily generated according to model (\ref{model_1}).
We call the program that implements variable selection according to
model (\ref{model_1}) \textbf{SEMMS} (Scalable EMpirical Bayes Model Selection), and
 compare its performance with several well-known variable selection approaches:
\begin{itemize}[noitemsep]
\item \textbf{\texttt{ncvreg}} \citep{Rncvreg}, which implements three types of penalties:
LASSO \citep{Tibshirani:1996}, SCAD \citep{FanLi:2001}, and MCP
\citep{Zhang:2010}.
\item \textbf{\texttt{glmnet}} \citep{glmnet}, to fit GLM models with the lasso or elastic net regularization.
\item \textbf{\texttt{lars}} (least angle regression) \citep{lars2013}.
\item \textbf{\texttt{SIS}} Sure Independence Screening \citep{SIS2018}. This package combines a
screening process which reduces the number of variables to be considered in each
regularization step, while ensuring that only irrelevant predictors are eliminated. This package
allows to choose among three penalty types (LASSO, SCAD, or MCP).
\item \textbf{\texttt{EMVS}} \citep{EMVS2018}, a Bayesian approach to variable selection with
a fast implementation via the EM algorithm.
\item \textbf{\texttt{spikeslab}} \citep{spikeslab2013}, a Bayesian (spike and slab) variable selection method.
\item \textbf{\texttt{mombf}} \citep{mombf2018} a Bayesian (MCMC) method based on \cite{johnson2012} using moment
and inverse moment Bayes Factors.
\item \textbf{\texttt{BoomSpikeSlab}} \citep{Scott2017}, a Bayesian (MCMC) implementation of the
spike and slab model.
\item \textbf{FDR} \citep{benjamini:hochberg} a simple one-predictor-at-a-time approach, controlling
the false discovery rate. In our simulations, we controlled the FDR at the 0.05 level.
\end{itemize}

The comparison with \texttt{spikeslab} and \texttt{EMVS} is only done in the normal
response case, since these packages do not have an option to fit a GLM model. Similarly,
\texttt{lars} only handles a normal linear regression model, but the package
\texttt{covTest} \citep{covTest2013} includes the function \texttt{lars.glm} which can also
be used to analyze binomial response and Cox regression models (but not Poisson).

Generally, these software packages were used with their default values. With \texttt{lars} and
\texttt{glmnet}, the number of selected variables was such that together they explained 90\% of the variability or (null) deviance.
The packages \texttt{ncvreg} and  \texttt{SIS} simply return the selected variables.
When using \texttt{spikeslab} we set the `bigp.smalln' option to TRUE, and
the selected variables were the non-zero generalized elastic net (gnet) coefficients.
With \texttt{EMVS} we varied the spike variance parameter (twenty values
between $10^{-10}$ and $10^{-1}$ equally spaced on the logarithmic scale)
and the type of the prior distribution
was set to `betabinomial'. The `independent' parameter was set to FALSE,
since it yielded much better results than the default (independent=TRUE, which
resulted in zero selected predictors in most cases.)
Hence, rather than running \texttt{EMVS} under the assumption that the regression
coefficients and the error variance are independent, a priori, a conjugate prior was used
\citep{Rockova:2014}.

In the case of a \textit{normal response} we show results from nine different scenarios with
varying dependence structures and number of significant predictors.
In each scenario $K=1000$ predictors were initially drawn independently from a $Unif[-1,1]$
distribution, but in scenarios 3, 4, and 5 they were modified in order to
induce correlations. Scenarios 1-5 are made under the assumption of our mixture model,
while 6-9 are not.
The true number of predictors in each scenario is denoted by L. The error terms, $\epsilon_i$,
are generated as i.i.d. $N(0,0.1)$ variates, except for N2 
where $\epsilon_i\sim N(0,0.25)$.

\begin{enumerate}[label=(N\arabic*)]
\item A single significant predictor ($L=1$) is related to the response:
$Y_i=Z_{1i}+\epsilon_i$.
\item The response is the sum of eight i.i.d. predictors ($L=8$):
 $Y_i=Z_{1i}+\ldots+Z_{8i}+\epsilon_i$.
\item $Y_i=Z_{1i}+\ldots+Z_{8i}+\epsilon_i$ (again, $L=8$)
but $Z_2=Z_1+\delta_{i2}$,  $Z_3=-2Z_1+\delta_{i3}$, $Z_4=-Z_1+\delta_{i4}$, and
$Z_6=-Z_5+\delta_{i6}$, where $\delta_{ik}\sim N(0,0.2)$, independently.
The correlation between $Z_1$ and each of $Z_2, Z_3, Z_4$ is approximately 0.95,
as is the correlation between $Z_5$ and $Z_6$.
\item $Y_i=Z_{1i}+\ldots+Z_{14i}+\epsilon_i$ (thus, $L=14$)
and $Z_2,\ldots,Z_{10}$ are drawn from a multivariate normal distribution with mean 0 and a
covariance matrix with a compound symmetry structure, $0.01\cdot I_9+ 0.05\cdot J_9$
where $I_9$ is a $9\times 9$ identity matrix and $J_9$ is a $9\times 9$ matrix of 1's.
The average correlation between any pair from the set $Z_2\,\ldots,Z_{10}$ is 0.93.
\item $Y_i=Z_{1i}+\ldots+Z_{20i}+\epsilon_i$ (thus, $L=20$)
where  $Z_1,\ldots, Z_{20}$ are drawn from a multivariate normal distribution
with an autoregressive (AR1) structure, with $\rho=0.95$.
\item $Y_i=\sum_{j=1}^{15}\beta_jZ_{ji}+\epsilon_i$ (thus, $L=15$) where $\beta_j\sim N(0,1)$, i.i.d.
In this case, the coefficients of the non-null predictors are drawn from the `slab' component in
the spike-and-slab model.
\item $Y_i=\sum_{j=1}^{15}\beta_jZ_{ji}+\epsilon_i$ (thus, $L=15$) where
$\beta_j=5, 1, 2, 4, 9, 3, 4, 1, 3, 2, 4, 2, 3, 1, 7$. In this case, the effect sizes
have dramatically different magnitudes.
\item  $Y_i=-\sum_{j=1}^{4}\beta_jZ_{ji}+\sum_{j=4}^{10}\beta_jZ_{ji}+\epsilon_i$ (thus, $L=10$) where $\beta_j=5, 7, 2, 4, 9, 3, 4, 1, 3, 2$.
Again, the magnitudes of effect sizes are very different, but this time
four of the effects are negative, and six are positive.
\item $Y_i=-\sum_{j=1}^{4}\beta_jZ_{ji}+\sum_{j=4}^{10}\beta_jZ_{ji}+\epsilon_i$ (thus, $L=10$) where $\beta_j=2, 2, 2, 2, 2, 2, 6, 6, 6, 6$.
Similar setting as the previous scenario in the sense that the magnitudes of effect sizes
are very different effects have positive and negative signs, but the effect sizes are from
two non-symmetric point masses (unlike our model, which assumes two normal distributions with
symmetric means.)
\end{enumerate}
In our simulations we used different sample sizes and different number of predictors. Results
shown here are for  $K=1,000$, and $N=100$  (Tables \ref{sim1} and \ref{sim3}) and $N=50$  (Table \ref{sim2}).
The median true positives and false positives were calculated from 30 replications of
each scenario. We observe several things:
\begin{itemize}[noitemsep]
\item For both values of $N$ SEMMS achieves the best or nearly the best results. When $N=100$ it finds all the
true predictors in simulations N1-N5 and N7-N8, and no false positive ones in any scenario.
When $N=50$, simulations N2-N4 yield a small number of
false positives, but the results are still very good.
\item In simulation N6 (Table \ref{sim3}) SEMMS finds 7 of the 15 true predictors, but keep in mind that
under the
spike and slab model which was used to generate the data, a large proportion of the slab component
overlaps with the spike component, and it is expected that under this model some true predictors will
not be detectable. Note that  \texttt{EMVS} which uses the spike and slab model for finding significant
predictors yields the same result. In this scenario \texttt{SIS} with a SCAD or MCP penalty achieves
a slightly better result.
\item When the effect sizes vary dramatically and do not follow our mixture prior assumption,
but are not concentrated at non-symmetric mass points (N6-N8), then SEMMS performs very well.
When the effect sizes are  concentrated at non-symmetric mass points (N9) SEMMS tends to find
the largest effect. This is perhaps an unlikely scenario, since we can expect the effect sizes to
vary, in which case SEMMS performs very well. In any case, because SEMMS maintains a very low
false positive rate, it is possible to deal with
cases such  as N9 by running SEMMS sequentially, each time moving the detected covariates to
the set of `locked-in' variables until no additional effects are found.
\item The spike and slab methods seem very sensitive to deviations from their assumed
model. In simulations N7-N9 both \texttt{EMVS} and \texttt{spikeslab} do not detect
all the true effects, and have a large number of false positives.
\item Perhaps a bit unexpected, in the $N=100$ case the one at a time approach (FDR)
performs quite well, especially in scenarios N1, N3, N4, and N5 where it is the second
or third best method in terms of
its overall error. However, FDR is quite conservative in simulations N6-N9, and
when $N$ is small relative to $K$ the power of this method is much lower
 in simulations N2-N4.
\item Between the \texttt{SIS} and \texttt{ncvreg} packages which offer similar regularization options
the former appears to be better at maintaining lower FP rate.
\item Between the two packages that offer a fast Bayesian (spike and slab) fitting, \texttt{EMVS} achieves
a lower FP rate in simulations N1-N4 and N6, but higher in simulation N5.
\item Generally, both \texttt{glmnet} and \texttt{lars} have high FP rates.
\end{itemize}

\begin{table}[H]
\centering
\caption{\label{sim1}Simulation study - normal response, $K=1,000$ predictors, $N=100$. $L$ is the
true number of predictors used in each scenario.}
\begin{tabular}{|l|rr|rr|rr|rr|rr|}
  \hline
  &\multicolumn{2}{c}{Sim. \#N1} &\multicolumn{2}{|c|}{Sim. \#N2} &\multicolumn{2}{|c|}{Sim. \#N3} &\multicolumn{2}{|c|}{Sim. \#N4} &\multicolumn{2}{|c|}{Sim. \#N5} \\
  &\multicolumn{2}{c}{$L=1$} &\multicolumn{2}{|c|}{$L=8$} &\multicolumn{2}{|c|}{$L=8$} &\multicolumn{2}{|c|}{$L=14$}
  &\multicolumn{2}{|c|}{$L=20$} \\
  Method & TP & FP & TP & FP & TP & FP & TP & FP & TP & FP \\
  \hline
\texttt{SEMMS} & 1 & 0 & 8 & 0 & 8 & 0 & 14 & 0 & 20 & 0  \\
\texttt{ncvreg} LASSO & 1 & 5 & 8 & 43.5 & 5 & 17 & 7 & 36 & 8 & 11.5 \\
\texttt{ncvreg} SCAD & 1 & 0 & 8 & 7 & 4 & 1.5 & 6 & 0 & 2 & 15.5  \\
\texttt{ncvreg} MCP & 1 & 0 & 8 & 0.5 & 4 & 0 & 6 & 0 & 2 & 7 \\
\texttt{glmnet} & 1 & 55 & 8 & 31 & 8 & 35 & 14 & 23 & 20 & 33 \\
\texttt{lars} & 1 & 41 & 8 & 21 & 8 & 19 & 11.5 & 0 & 20 & 25.5 \\
\texttt{SIS} LASSO & 1 & 0 & 8 & 8 & 5 & 1 & 7 & 1 & 16 & 5 \\
\texttt{SIS} SCAD & 1 & 0 & 8 & 0 & 4 & 0 & 6 & 0 & 13 & 8 \\
\texttt{SIS} MCP & 1 & 0 & 8 & 0 & 4 & 0 & 6 & 0 & 12.5 & 8.5 \\
\texttt{EMVS} & 1 & 0 & 8 & 0 & 5 & 0 & 6 & 0 & 16.5 & 39 \\
\texttt{spikeslab} & 1 & 0 & 8 & 16 & 6 & 11 & 6 & 5 & 11 & 14  \\
FDR & 1 & 0 & 2.5 & 0 & 7 & 0.5 & 9 & 0 & 20 & 1 \\
   \hline
\end{tabular}
\end{table}

\begin{table}[H]
\centering
\caption{\label{sim3}Simulation study - normal response, $K=1,000$ predictors, $N=100$. $L$ is the
true number of predictors used in each scenario. In these simulations the true distribution of
the significant effects is different from our mixture model.}
\begin{tabular}{|l|rr|rr|rr|rr|}
  \hline
  &\multicolumn{2}{c}{Sim. \#N6} &\multicolumn{2}{|c|}{Sim. \#N7} &\multicolumn{2}{|c|}{Sim. \#N8} &\multicolumn{2}{|c|}{Sim. \#N9} \\
  &\multicolumn{2}{c}{$L=15$} &\multicolumn{2}{|c|}{$L=15$} &\multicolumn{2}{|c|}{$L=10$} &\multicolumn{2}{|c|}{$L=10$} \\
  Method & TP & FP & TP & FP & TP & FP & TP & FP \\
  \hline
\texttt{SEMMS}  & 7 & 0 & 15 & 0 & 10 & 0 & 4 & 0\\
\texttt{ncvreg} LASSO & 11 & 30.5 & 15 & 17.5 & 10 & 3 & 10 & 3\\
\texttt{ncvreg} SCAD  &12 & 11.5 & 15 & 0 & 10 & 0 & 10 & 0\\
\texttt{ncvreg} MCP & 11 & 5& 15 & 0 & 10 & 0 & 10 & 0\\
\texttt{glmnet} & 15 & 85 & 15 & 27 & 10 & 30  & 10 & 27\\
\texttt{lars} & 15 & 139 & 15 & 4.5 &  10 & 0 & 10 & 1.5\\
\texttt{SIS} LASSO & 8.5 & 3 & 10 & 11 & 10 & 4 & 10 & 5\\
\texttt{SIS} SCAD & 8.5 & 0 & 15 & 0 & 10 & 0 & 10 & 0\\
\texttt{SIS} MCP & 8 & 0 & 15 & 0 & 10 & 0 & 10 & 0\\
\texttt{EMVS} & 7 & 0 & 9 & 43 & 7 & 40.5 & 7 & 36 \\
\texttt{spikeslab}  & 6 & 12.5 & 9 & 14 & 7 & 35 & 7 & 39\\
FDR  & 2 & 0 & 3 & 0 & 3 & 0 & 4 & 0\\
   \hline
\end{tabular}
\end{table}

\begin{table}[H]
\centering
\caption{\label{sim2}Simulation study - normal response, $K=1,000$ predictors, $N=50$. $L$ is the
true number of predictors used in each scenario.}
\begin{tabular}{|l|rr|rr|rr|rr|rr|}
  \hline
  &\multicolumn{2}{c}{Sim. \#N1} &\multicolumn{2}{|c|}{Sim. \#N2} &\multicolumn{2}{|c|}{Sim. \#N3} &\multicolumn{2}{|c|}{Sim. \#N4} &\multicolumn{2}{|c|}{Sim. \#N5}\\
  &\multicolumn{2}{c}{$L=1$} &\multicolumn{2}{|c|}{$L=8$} &\multicolumn{2}{|c|}{$L=8$} &\multicolumn{2}{|c|}{$L=14$} &\multicolumn{2}{|c|}{$L=20$}\\
 Method & TP & FP & TP & FP & TP & FP & TP & FP & TP & FP \\
  \hline
\texttt{SEMMS} & 1 & 0 & 2 & 4 & 8 & 2 & 7 & 2.5 & 20 & 0 \\
\texttt{ncvreg} LASSO & 1 & 8.5 & 5.5 & 25 & 5 & 24 & 7 & 30 & 6 & 13.5 \\
\texttt{ncvreg} SCAD & 1 & 0 & 4 & 12.5 & 4 & 11.5 & 6 & 3.5 & 2 & 5 \\
\texttt{ncvreg} MCP & 1 & 0 & 1.5 & 2 & 4 & 4 & 6 & 1 & 1 & 1.5 \\
\texttt{glmnet} & 1 & 45 & 8 & 27 & 8 & 29 & 14 & 21 & 20 & 24 \\
\texttt{lars} & 1 & 22 & 8 & 25 & 8 & 13 & 14 & 10.5 & 20 & 5.5 \\
\texttt{SIS} LASSO & 1 & 0 & 3 & 9 & 4 & 4 & 5.5 & 6 & 11 & 1 \\
\texttt{SIS} SCAD & 1 & 0 & 3 & 9 & 4 & 1 & 6 & 0 & 9 & 3 \\
\texttt{SIS} MCP & 1 & 0 & 2.5 & 8 & 4 & 0 & 6 & 0 & 9.5 & 2.5 \\
\texttt{EMVS} & 1 & 0 & 3 & 5.5 & 5 & 0 & 5 & 0.5 & 16 & 26.5 \\
\texttt{spikeslab} & 0 & 0 & 4 & 25 & 4 & 8.5 & 6 & 13.5 & 9 & 8 \\
FDR & 1 & 0 & 0 & 0 & 2 & 0 & 1 & 0 & 20 & 1.5 \\
   \hline
\end{tabular}
\end{table}

We obtained similar results in the \textit{binary response} case.
In Table \ref{sim4} we show results from two scenarios, both with $K=1000$ and $N=120$.
We only use the methods which perform variable selection in the GLM framework,
namely, \texttt{ncvreg}, \texttt{glmnet},  \texttt{SIS}, and the one-at-a-time approach,
controlling the FDR. The two simulation scenarios are:
\begin{enumerate}[label=(B\arabic*)]
\item $\eta_i=2Z_{3i}+2Z_{6i}+2Z_{7i}$
but $Z_2=Z_1+\delta_{i2}$,  $Z_3=-2Z_1+\delta_{i3}$, $Z_4=-Z_1+\delta_{i4}$, and
$Z_6=-Z_5+\delta_{i6}$, where $\delta_{ik}\sim N(0,0.2)$, independently.
The correlation between $Z_1$ and each of $Z_2, Z_3, Z_4$ is approximately 0.95,
as is the correlation between $Z_5$ and $Z_6$. Thus, in this case $L=7$.
\item Autoregressive structure: we set $L=10$ and  $\eta_i=2Z_{1,i}+2Z_{101,i}$
but $Z_1-Z_5$ have an AR(1) structure with $\rho=0.95$ and so do  $Z_{101}-Z_{105}$.
\item Hub network - we create an $N\times K$ matrix so that the columns consist of
$g$ non-overlapping hubs such that within each
hub all $K/g$ nodes follow a multivariate normal distribution with compound symmetry correlation structure.
We pick one node in one of the hubs to be the response, and replace it with a vector of $N$ 0/1 values, based on
a logistic model with the other $L=(K/g)-1$ nodes in that hub as predictors. In Table \ref{sim4}
we show results when $L=9$ (i.e. 100 hubs, each with 10 nodes), but we get very similar with $L=4$ and $L=19$.
For this configuration Table \ref{sim4} includes results with two sample sizes: $N=120$ and $N=80$.
\end{enumerate}
As was the case with the normal response,
SEMMS appears to
have the best overall performance with the binary responses data. FDR has a slightly higher power in simulation B3 when $N=120$, but it
detects zero variables when $N=80$, whereas the performance of SEMMS remains approximately the same
when $N$ is decreased from 120 to 80.

Similarly to the normal and binomial responses, our method has the lowest median false positive rate
and a high power to detect the true effects in the \textit{Poisson} model. Table \ref{sim4} shows the results
of simulation (P1), where $\eta=3+0.3Z_1 + 0.25Z_2 -0.22Z_3 -0.19Z_4 + 0.27Z_5 -0.17Z_6 -0.25Z_7$
and $N=120$.
Simulation (P2) uses the same model, except that predictors $Z_1-Z_5$ have an AR(1) structure with $\rho=0.95$.
With P1, SEMMS has a median TP five predictors out of seven, and median of 0 false positives.
In  the P2 setting SEMMS detects all true predictors, while still having no false positives.
The competing methods \texttt{ncvreg}, \texttt{glmnet},  \texttt{SIS} all achieve good
results in terms of power in P1, detecting all seven predictors (and fewer in P2), but a higher false positive count, as compared with
SEMMS. Among \texttt{ncvreg}, \texttt{glmnet}, and \texttt{SIS} the latter appears to yield lower false positive rates, yielding a median of five with all three penalty types in P1. In P2 \texttt{SIS MCP} has a median of 0.5 false positive,
but it detects only three of the seven true predictors.
Note that in the one-at-a-time approach, we used the \texttt{family=quasipoisson()} option, and not
 \texttt{family=poisson()} since the latter yielded a very high false positive rate.

\begin{table}[t!]
\centering
\caption{\label{sim4}Simulation study - binary response (simulations B1 and B2)
and the Poisson model (simulation P1), all with $K=1,000$ predictors, $N=120$.
For Sim. \#B3 we also show results for $N=80$. $L$ is the
true number of predictors used in each scenario. }
\begin{tabular}{|l|rr|rr|rr|rr|rr|rr|}
  \hline
  &\multicolumn{2}{c}{Sim. \#B1} &\multicolumn{2}{|c|}{Sim. \#B2} &\multicolumn{4}{|c|}{Sim. \#B3}  &\multicolumn{2}{|c|}{Sim. \#P1} &\multicolumn{2}{|c|}{Sim. \#P2}\\
  &\multicolumn{2}{c}{$L=7$} &\multicolumn{2}{|c|}{$L=10$} &\multicolumn{4}{|c|}{$L=9$ \footnotesize{($N=120,80$)}} &\multicolumn{2}{|c|}{$L=7$} &\multicolumn{2}{|c|}{$L=7$} \\
 Method & TP & FP & TP & FP & TP & FP & TP & FP & TP & FP & TP & FP\\
  \hline
\texttt{SEMMS}        & 6 & 2 & 7.5 & 3.5 & 5 & 0 & 4 & 1 & 5 & 0 & 7 & 0\\
\texttt{ncvreg} LASSO & 3.5 & 12.5 & 1.5 & 4.5 & 7 & 23 & 6 & 14 & 7 & 37.5 & 4 & 21.5\\
\texttt{ncvreg} SCAD  & 3 & 12 & 1.5 & 4.5 & 7 & 18 & 5 & 12.5 & 7 & 28 & 3 & 17\\
\texttt{ncvreg} MCP   & 3 & 5.5 & 1 & 0.5 & 5 & 5 & 3 & 3.5 & 7 & 8.5 & 3 & 3.5\\
\texttt{glmnet}       & 7 & 58.5 & 5 & 43.5 & 9 & 53 & 9 & 48 & 7 & 40 & 7 & 47\\
\texttt{SIS} LASSO    & 3 & 2 & 2 & 4 & 4 & 2 & 2 & 2 & 7 & 5 & 5 & 6\\
\texttt{SIS} SCAD     & 3 & 3 & 1 & 4 & 4 & 2 & 2 & 2 & 7 & 5 & 3 & 6.5 \\
\texttt{SIS} MCP      & 3 & 3 & 1 & 4 & 4 & 2 & 2 & 2 & 7 & 5 & 3 & 0.5\\
FDR                   & 4 & 0 & 0 & 0 & 6 & 0 & 0 & 0 & 3 & 0 & 6 & 0\\
   \hline
\end{tabular}
\end{table}

In our simulation results we did not include the two MCMC-based methods
 (\texttt{mombf} and \texttt{BoomSpikeSlab}) because they proved to be far too slow to be
 practical, given the existence of
fast alternatives, including Bayesian approaches like \texttt{spikeslab} and especially \texttt{EMVS}.
For example, when we ran simulations N1 and N5 with \texttt{mombf} but with a much smaller number of predictors
($K=100$) scenario N1 took an average of 4 minutes to complete each replicate, and N5 took an
average of 1.6 hours
(with a sample size $N=30$ and 1,000 MCMC iterations in both scenarios.)
In the N1 case, the correct model was detected almost every time, but in the case of N5
the median number of true positives was 4 (when $L$ is actually 20), and the median FP was 0.
It appears that \texttt{mombf} gets much slower as $L$ increases, and less powerful
as compared with competing methods.

The \texttt{BoomSpikeSlab} package appears to be somewhat faster, taking about 40 minutes to
complete 1,000 MCMC iterations when $K=1000$. However, it does not seem to have an advantage
over the best available methods in terms
of power or error rate. For example, in the case of simulation N5
the median number of predictors found by \texttt{BoomSpikeSlab} was 1 (out of 20), with
0 false positives.

\section{Case Studies}\label{sec:vs:cases}
\subsection{Normal Response -- the Riboflavin Data}
In a recent paper demonstrating modern approaches to high-dimensional statistics,
\cite{buhlmann2014} analyzed a data set from \cite{lee:2001} in which the response
variable is the logarithm of riboflavin (vitamin B12) production rate,
and there are normalized expression levels of 4,088 genes which are used
as explanatory variables. The sample size in the data set is $N=71$.
In addition to the fact that the number of putative variable greatly exceeds
the number of observations, many of the putative variables are highly correlated.
Out of 8,353,828 pairs of genes, there are 70,349 with correlation coefficient
greater than 0.8 (in absolute value).

\cite{buhlmann2014} report that the Lasso with $B=500$ independent random subsamples of size $\lfloor N/2 \rfloor$,
and with $q=20$ variables that enter the regularization step
first, yields three significant and stable genes: LYSC\_at, YOAB\_at, and YXLD\_at.
The model with these three variables has an $R^2$ of 0.68, and AIC of 118.6.
Their multisample-split method yields one significant variable (YXLD\_at), and the
projection  estimator, used with Ridge-type score yields no significant variables
at the FWER-adjusted 5\% significance level.
The model with just YXLD\_at has an $R^2$ of 0.37 and AIC of 162.

\cite{LedererM15} also used this data set, to demonstrate their TREX model. Their final model
includes three genes: YXLE\_at, YOAB\_at, and YXLD\_at. The $R^2$ of their
model is 0.62, and the AIC is 130.68. Two of the genes are highly correlated (YXLD\_at and
YXLE\_at) and yield a variance inflation factor of 23.7 each, so when fitting the final
linear regression model neither appears to be significant.

\begin{figure}[t!]
\begin{center}
\centerline{\includegraphics[trim=0mm 0mm 0mm 10mm,clip,width=6in]{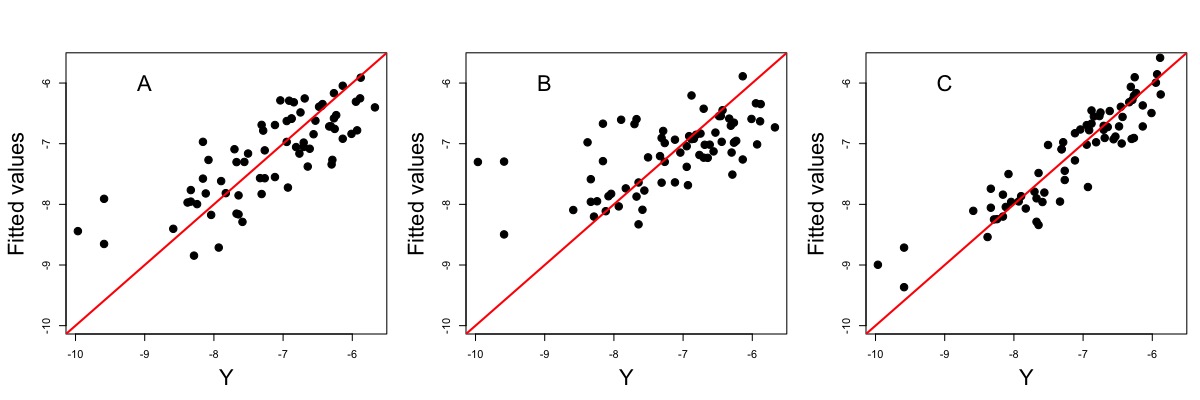}}
\caption{Riboflavin data -- fitted vs. observed values.  A: \cite{LedererM15},
B: \cite{buhlmann2014}, C: SEMMS - greedy algorithm.
}
\label{riboflavin.aic}
\end{center}
\end{figure}

We ran SEMMS using both the greedy and the weighted probability methods as
described in Section \ref{sec:gam}.
The greedy method yielded six significant genes: LYSC\_at, SPOIISA\_at , XTRA\_at,
YDDK\_at, YURQ\_at, and YXLD\_at. The predicted level of the logarithm of
riboflavin production rate is given by the formula:
\begin{eqnarray*}
\hat{Y}&=&-7.16 - 0.23\cdot Z_{LYSC\_at} + 0.19\cdot Z_{SPOIISA\_at}
+ 0.16\cdot Z_{XTRA\_at} \\ &&
- 0.23\cdot Z_{YDDK\_at} + 0.26\cdot Z_{YURQ\_at}
-0.35\cdot Z_{YXLD\_at}
\end{eqnarray*}
This model has AIC=70.9  and an $R^2$ of 0.85.
Figure \ref{riboflavin.aic}  depicts the observed values
 (Y) vs. the fitted values from three models (\cite{buhlmann2014}, \cite{LedererM15},
 and greedy SEMMS.)
In addition to having much smaller residuals than the two other methods,
SEMMS provides much better prediction for low values of riboflavin. The other two
methods seem to  over-estimate the riboflavin levels when the true (normalized) values
are small (less than $-9$).

We ran SEMMS using the weighted probability method 100 times.
The best model included six genes (CARB\_at, SPOVAB\_at, XHLA\_at,
YCKE\_at, YOAB\_at, YXLD\_at) and had an AIC of  53.9 and an
$R^2$ of 0.88.
In the 100 runs of the randomized SEMMS a total of 16 genes were selected, yielding
an AIC of 46.8 and an $R^2$ of 0.92.

We also used packages that performed well in our simulations. \texttt{EMVS} found
eight predictors (AIC=68.2) and \texttt{SIS} with MCP penalty found four predictors
(AIC=86.3).

\begin{figure}[t!]
\begin{center}
\centerline{\includegraphics[trim=10mm 170mm 10mm 20mm,clip,width=6in]{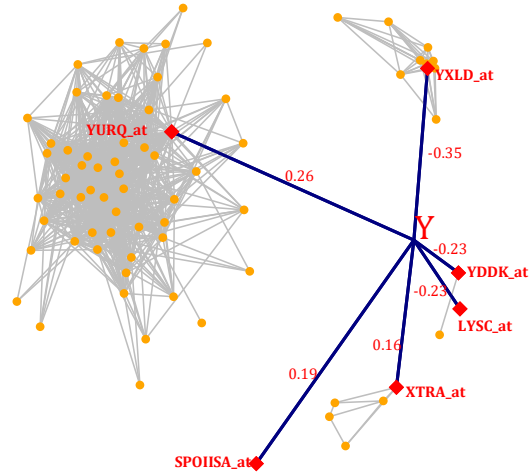}}
\caption{Riboflavin data -- a graphical representation of the model found by SEMMS using the greedy
algorithm.}
\label{riboflavin.network}
\end{center}
\end{figure}

The notion of `\textit{the best} selected model' may not always be appropriate, since
(i) the number of putative variables is large there is no way to evaluate all possible
models, and (ii) some selected predictors can be part of a network of highly correlated
variables.
To deal with (ii) SEMMS detects predictors that are highly correlated with ones
selected to be in the model, and reports them as well. This is illustrated in
Figure \ref{riboflavin.network}.
The variables selected by (greedy) SEMMS are shown as red diamonds and their
 coefficients in the fitted linear model as numbers next to the (dark blue) edges.
Variables that are highly correlated with a selected predictor are depicted as orange dots and
they are connected via a gray edge to another predictor if their correlation with that predictor
exceeds a user-defined threshold (in this case, we used 0.75).
With the riboflavin dataset it can be seen that the YURQ\_at gene is co-expressed with a
large group of genes, and hence each one of those genes (or perhaps some
weighted average of their expression levels) could be considered as a relevant predictor
for the response. In this case study, with a correlation threshold of 0.75 between predictors
greedy SEMMS reports a total of 62 relevant variables.

\subsection{Binary Response -- the BMI Data}
\cite{Lin13082014} demonstrated an application of a LASSO-based variable selection
method for regression models with compositional covariates. The analysis aims
to identify a subset of 87 bacteria genera in the gut whose subcomposition
is associated with body-mass index (BMI). The data, which was introduced in
\cite{wu:2011}, is compositional,
which using our previous notation means that $\sum_{j=1}^{87} z_{ij}=1$ for each $i$.
The total number of samples is $N=96$.
To apply our method directly, without changing the model to account for the sum
constraint, we perform the log ratio transformation and replace the matrix $\mathbf{Z}$
with $\mathbf{Z}^K=[\log(z_{ij}/z_{iK})]$. The data contains many zero counts, so
\cite{Lin13082014} replace them with 0.5 before converting the data to be in
compositional form. Note that there may be preferable imputation methods, but we choose
to use the same method in \cite{Lin13082014} in order to have a meaningful comparison.
We use a subset of
45 bacteria which had non-zero counts in at least 10\% of the samples ($N=96$).  The
omitted genera have minimal contribution to the overall distribution of
the proportions.

To demonstrate the application of SEMMS to the binary response case we create
an categorical BMI variable with 4 levels: underweight - less than 18.5 (n=5),
normal [18.5, 25) (n=30), overweight [25.5, 30) (n=25), and obese  $\ge$30 (n=10).
We used the binomial option in SEMMS
in order to find  which bacteria are associated with $p_b=Pr(obese)$ and
which are associated with $p_w=Pr(overweight)$, with the normal group as the baseline.
The underweight group is too small and is not included in the analysis.
We found six bacteria associated with obesity
(\emph{Acidaminococcus, Alistipes, Allisonella, Butyricimonas, Clostridium,
and Oxalobacter}), and five with overweight
(\emph{Anaerofilum, Faecalibacterium, Oscillibacter, Turicibacter, and Veillonella}).
We also ran SEMMS with BMI as a continuous response, using the normal model and
obtained the same four genera reported by \cite{Lin13082014}
(\emph{Acidaminococcus, Alistipes,  Allisonella, and Clostridium}).
The fact that these four are a subset of the genera found to be associated with
$p_b$ but have no overlap with the ones associated with $p_w$
suggest that different BMI levels are  associated with different
bacteria, and thus, a categorical analysis or perhaps quantile regression
may be more appropriate than the conditional mean models in regression analysis.

\subsection{Survival Analysis -- the NKI70 data}
Our model can be used to deal with censored survival times.
We follow \citet{Whitehead:1980} who proposed using an
artificial Poisson model to fit Cox's proportional hazards (PH) regression model.
Recall that in the proportional hazard model, the hazard functions have the form
$\lambda(t;\mathbf{z})=\lambda_0(t)\exp(\mathbf{z}\bm\beta)$.
Suppose there are $q$ deaths occurring after survival times $t_1,\ldots,t_q$
(one death after each $t_i$.)
For each $h=1,\ldots,q$ let $Y_{h,j}$ be independent Poisson random variables with
parameters $\mu_{h,j}$, such that $$\mu_{h,j}=N_{h,j}\exp(\alpha_h+\mathbf{z}_{h,j}\bm\beta)$$
where $N_{h,j}$ is the number of patients who are at risk after $t_h$
and whose explanatory vector is $\mathbf{z}_{h,j}(t_h)$.
Since the excess life of a patient beyond time $t_h-\delta t_h$ follows an exponential
distribution, the number of deaths in each group follows a Poisson distribution
with parameter $\mu_{h,j}$.
In other words, rather than modeling the survival times,
we model the number of (instantaneous) deaths among the survivors at a given time
in a given group.
The mean number of instantaneous deaths among survivors in a group of
individuals sharing the same properties is associated with a linear combination of
the explanatory variables through the log link function.
This strategy allows for ties and censoring in the data.
Another approach to survival modeling is the  accelerated failure time (AFT) models.
More biostatisticians are using the AFT model because it is based on the linear model
and the estimated regression coefficients have a rather direct physical interpretation.
The AFT model can be formulated as variants of estimating equations that
correspond to generalized linear models via a Buckley-James transformed Gaussian estimating equations
\citep{buck:jame:1979}.

To demonstrate this approach, we used the NKI70 dataset from \cite{nki70:2002}, which is
available in the `\texttt{penalized}' package \citep{penalized:2010}.
This dataset contains gene expression measurements for 70 genes,
obtained from lymph nodes of 144 breast cancer patients.
The 70 genes were determined as prognostic for metastasis-free survival
in earlier studies.  Large-scale microarray gene expression analyses have revealed
a signature set of genes that can predict breast-cancer prognosis
(\cite{nki70:2002}, \cite{van2002gene}, \cite{wang2005}, \cite{fan2006}, \cite{hua2008}, \cite{madden2013}).
However, few genes overlapped in these assays, and only a few of
the breast-associated genes have been validated at the protein
level.

We analyzed which genes were associated with either death or recurrence
of metastasis in less than three months. We included the (scaled) age of
the patients as an explanatory variable, as well as the logarithm of the
time until death/recurrence or censoring where, for the purpose of
this analysis we considered any
individual who survived more than 3 months as censored.  SEMMS generated the following model
\begin{eqnarray*}
 \eta=\log\mu=-2.8-0.37\log(t)-0.35\cdot Z_{SCUBE2} - 0.79\cdot Z_{KNTC2}\\
 -0.56\cdot Z_{ZNF533}+0.36\cdot Z_{IGFBP5.1}+1.09\cdot Z_{PRC1}
\end{eqnarray*}
The residual deviance for this model is 39 (vs. 87.55 for an intercept-only model).
These selected genes have been  determined to be critical  as prognostic for metastasis-free survival
in an earlier studies.
In particular, a cross-platform comparison of breast-cancer
gene sets from these profiling studies revealed SCUBE2 as a common gene that has been validated at the protein
level  \citep{lin2014}.

%

\section{Conclusion}\label{sec:varsel:conc}
In 1996 Brad Efron stated that variable selection in regression is the most important
problem in statistics \citep{Hesterberg:2010}. Since then many papers have been written
on the topic, as this continues to be a challenging problem in the age of high-throughput
sequencing in genomics, and as other types of `omics' data become available and more
affordable.
We have developed a model-based, empirical Bayes approach to variable
selection. We define a mixture model in which the putative
variables are modeled as random effects, and we demonstrate that this approach
results in high power to select the correct variables
while maintaining a low rate of false positive selections, in a variety
of situations.
Our algorithm is scalable and computationally efficient because of: (i) the
parsimony of the mixture model, since
the mixture model involves a very small number of parameters, and that number
remains constant regardless of the number
of putative variables; and (ii) the usage of the Generalized Alternating
Maximization algorithm to estimate the model parameters, as well as
an efficient dimension reduction trick via the Woodbury identity.
The Generalized Alternating Maximization algorithm not only converges
significantly faster than simulation-based methods, but also uses
memory more efficiently, since it has to keep only the posterior non/nonnull
probabilities for each predictor from only the two latest iterations,
as opposed to entire chains.
Finally, a simple modification to the algorithm prevents multicollinearity problems in
the fitted regression model.

\bigskip
\begin{center}
{\large\bf SUPPLEMENTARY MATERIAL}
\end{center}

\begin{description}

\item[Title:] The file SEMMS\_a.pdf contains some implementation notes, instructions how
to obtain and install the package, and three examples with code and plots generated by
SEMMS. (pdf format)

\item[R-package:] version 0.1.0 of  SEMMS is available online at \url{https://haim-bar.uconn.edu/software/}.
The GNU zipped tar file contains datasets used in the case studies, as well as one
simulated dataset and a file with ozone levels data for an additional case study of a well-studied
dataset.

\end{description}

\appendix
\section{Appendix}\label{proof}
{\bf Proof of Proposition \ref{convergencethm}}:
We denote the estimates obtained in step $t$ of Algorithm \ref{alg1} by a superscript.
First, we show that the algorithm terminates after a finite number of steps.
Given $\bm\gamma^t$, the estimate for $\bm\theta$ from Line \ref{maxtheta} is unique, because the log-likelihood
is a convex function. Therefore, if at any $t$, $S^{t+1}$ is empty (that is, the loglikelihood will not improve
by changing $\bm\gamma$) then
the algorithm terminates because the estimate for $\bm\theta$ will not change in
the subsequent iteration.
So, we need to show that for all $t\ge 0$ and $s>1$ we cannot have
$\bm\gamma^{t}=\bm\gamma^{t+s}$ (i.e., there are no loops in the sequence of $\bm\gamma^{t}$).
Suppose that for some $t$ there exists $s>1$ such that $\bm\gamma^{t}=\bm\gamma^{t+s}$. Then,
the convexity of the loglikelihood function implies that $\bm\theta^{t}=\bm\theta^{t+s}$. Together, this means that
$\ell^{t}(\mathbf{y}) = \ell^{t+s}(\mathbf{y})$, but in the algorithm we selected
the coordinates to modify in each step, $k_t,\ldots,k_{t+s}$, so that in each
iteration the log-likelihood strictly increased, so we must have
$\ell^{t+s}(\mathbf{y}) - \ell^{t}(\mathbf{y}) > 0$.
Proposition 7 in \cite{Gunawardana:2005} shows that minimization (and
consequently maximization) of a continuous function forms a closed
point-to-set map.  Their Lemma 9 shows the Cartesian product of
two closed point-to-set-maps is itself closed.  With these results in
hand one can apply Zangwill's generalized convergence theorem \citep{Zangwill1969} to
the sequence $(\bm\theta^t,  \bm\gamma^t)$.  \QED

\bibliographystyle{chicago}
\bibliography{refs}
\end{document}